\documentclass[aps,preprint]{revtex4}
\usepackage{amssymb,epsf}
\usepackage{amsmath}

\begin{document}

\title{Thermodynamics of Rotating Lovelock-Lifshitz Black Branes}
\author{M. H. Dehghani$^{1,2}$\footnote{mhd@shirazu.ac.ir} and Sh. Asnafi$^1$}
\affiliation{$^1$ Physics Department and Biruni Observatory,
College of Sciences, Shiraz University, Shiraz 71454, Iran\\
$^2$ Research Institute for Astrophysics and Astronomy of Maragha
(RIAAM), Maragha, Iran}

\begin{abstract}
We investigate the thermodynamics of rotating Lovelock-Lifshitz black branes.
We calculate the conserved and thermodynamic quantities of the solutions and
obtain a relation between energy, entropy and angular momentum densities with temperature and angular velocity.
We also obtain a Smarr-type formula for the energy density as a function of entropy and angular momentum densities,
and we show that the thermodynamic quantities calculated in this paper satisfy
the first law of thermodynamics. Finally, we investigate the stability of
black brane solutions in both canonical and grand-canonical ensemble. We
find that the solutions are thermally stable for $z\leq n-1$, while they can be unstable for $z>n-1$.
\end{abstract}
\maketitle

\section{Introduction}

Holography presents a general tool for studying many kinds of physical
systems. It is known that finite temperature in the field theory often
corresponds to the presence of a horizon in its gravitational dual. Although
the idea of holography has been first used in the AdS/CFT correspondence,
recently these techniques have been brought to bear on other types of
systems \cite{KBal}. Most notably, the holographic techniques are also useful to study
condensed matter systems at strong coupling \cite{Condens}. Indeed, the
Lifshitz black holes have been found to emerge as gravity duals of some
condensed matter systems with anisotropic scaling symmetry,
\[
t\rightarrow \lambda ^{z}t,\qquad \mathbf{x}\rightarrow \lambda \mathbf{x},
\]
where $z$ is the dynamical exponent.

From a holographic point of view, this suggests the following asymptotic
form for the spacetime metric
\begin{equation}
ds^{2}=-\frac{r^{2z}}{l^{2z}}dt^{2}+\frac{l^{2}}{r^{2}}dr^{2}+r^{2}d\varphi
^{2}+r^{2}\sum_{i=1}^{n-2}dx_{i}^{2},  \label{Stat}
\end{equation}
which is known as Lifshitz spacetime and obeys the scale invariance
\[
t\rightarrow \lambda ^{z}t,\qquad r\rightarrow \lambda ^{-1}r,\qquad \mathbf{%
x}\rightarrow \lambda \mathbf{x}.
\]
The metric (\ref{Stat}) is the well-known AdS metric for $z=1$. This spacetime possesses a timelike Killing vector, but is not a space of
constant curvature. In order to have solutions with Lifshitz asymptotic, one
may use an action involving a 2-form and a 3-form field with a
Chern-Simons coupling or a massive vector field \cite{Ka}. Asymptotic
Lifshitz solutions has been investigated by many authors. Only a few exact
solutions have been found, and most of the solutions have been obtained
numerically \cite{Lif}.

Because of the fact that corrections from higher powers of the curvature must be
considered on the gravity side of the correspondence in order to investigate
CFTs with different values of their central charges, the holography of
gravity theories including higher powers of the curvature have attracted
increased attention \cite{Boer, Myers}. Asymptotic Lifshitz solutions in the
vacuum of higher-derivative gravity have been investigated and it is shown
that the higher-curvature terms with suitable coupling constant may play the
role of the desired matter \cite{HDG}. Recently, one of us introduced some
exact and numerical Lifshitz solutions in higher-curvature gravity with
cubic-curvature terms \cite{Deh0,BDM}. From the point of view of holography,
thermodynamics of Lifshitz black holes has been investigated by many authors
in different theories of gravity \cite{Therm}.

In this paper, we would like to investigate the thermodynamics of rotating Lifshitz black branes
in third-order Lovelock gravity. We use the counterterm method introduced in
\cite{Deh1} in order to compute the conserved quantities of the spacetime.
The motivation for considering rotating Lifshitz black branes is to investigate the effect of
rotation parameter on the properties of black branes. Specially, we would like to find the effect of rotation parameter
on the stability of rotating Lifshitz black branes. Although asymptotically AdS black branes
with flat horizon are stable \cite{DKh,Witten}, we find that asymptotic Lifshitz
black branes with a flat horizon can be unstable.

The outline of our paper is as follows. In Sec. \ref{Gen}, we
give a brief review of general formalism of
calculating the conserved quantities of Lifshitz black branes. In Sec. \ref{Rot},
we introduce rotating Lifshitz black branes. Section \ref{Therm} is devoted
to the investigation of the thermodynamics of rotating Lifshitz black branes.
We also perform a local stability analysis of the black branes in the canonical and grand-canonical ensembles.
We finish our paper with some concluding
remarks.
\section{General Formalism\label{Gen}}

Here, we give a brief review of the general formalism of calculating the
conserved quantities of asymptotically Lifshitz black branes in third-order
Lovelock gravity. The action of third-order Lovelock gravity may be written
as
\begin{equation}
I_{\mathrm{g}} =\frac{1}{16\pi }\int d^{n+1}x\sqrt{-g}\left(
\sum_{i=1}^{[n/2]}\alpha _{i}\mathcal{L}_{i}-2\Lambda -\frac{1}{4}F_{\mu \nu
}F^{\mu \nu }-\frac{1}{2}m^{2}A_{\mu }A^{\mu }\right),   \label{action}
\end{equation}
where $[x]$ is the integer part of $x$, $\Lambda $ is the cosmological
constant, and $\alpha _{i}$'s are Lovelock coefficients with $\alpha _{1}=1$. In
the above action, $F_{\mu \nu }=\partial _{\lbrack \mu }A_{\nu ]}$ where $%
A_{\mu }$ representing a massive gauge field with mass $m$, and $\mathcal{L}%
_{i}$ is the $i$th order Lovelock Lagrangian given as
\begin{equation}
\mathcal{L}_{i}=\frac{1}{2^{i}}\delta _{\nu _{1}\nu _{2}\cdots \nu
_{2i}}^{\mu _{1}\mu _{2}\cdots \mu _{2i}}R_{\mu _{1}\mu _{2}}^{%
\phantom{\alpha_1\alpha_2}{\nu_1\nu_2}}\cdots R_{\mu _{2i-1}\mu _{2i}}^{%
\phantom{\alpha_{2i-1} \alpha_{2i}}{\nu_{2i-1} \nu_{2i}}}.  \label{Act1}
\end{equation}
In Lovelock gravity only terms with order less than or equal $[n/2]$ contribute
to the field equations, the rest being total derivatives in the action.
Here, we consider Lovelock gravity up to a third-order term and therefore we
consider $(n+1)$-dimensional spacetimes with $n\geq 6$ (though in situations
where we set $\alpha _{2}=\alpha _{3}=0$ our solutions will be valid for $%
n\geq 2$). The explicit form of $i$th order Lovelock Lagrangian up to the third-order are $%
\mathcal{L}_{1}=R$, $\mathcal{L}_{2}=R_{\mu \nu \gamma \delta }R^{\mu \nu
\gamma \delta }-4R_{\mu \nu }R^{\mu \nu }+R^{2}$ and

\begin{eqnarray}
\mathcal{L}_{3} &=&R^{3}+2R^{\mu \nu \sigma \kappa }R_{\sigma \kappa \rho
\tau }R_{\phantom{\rho \tau }{\mu \nu }}^{\rho \tau }+8R_{\phantom{\mu
\nu}{\sigma \rho}}^{\mu \nu }R_{\phantom {\sigma \kappa} {\nu \tau}}^{\sigma
\kappa }R_{\phantom{\rho \tau}{ \mu \kappa}}^{\rho \tau }+24R^{\mu \nu
\sigma \kappa }R_{\sigma \kappa \nu \rho }R_{\phantom{\rho}{\mu}}^{\rho }
\nonumber \\
&&+3RR^{\mu \nu \sigma \kappa }R_{\sigma \kappa \mu \nu }+24R^{\mu \nu
\sigma \kappa }R_{\sigma \mu }R_{\kappa \nu }+16R^{\mu \nu }R_{\nu \sigma
}R_{\phantom{\sigma}{\mu}}^{\sigma }-12RR^{\mu \nu }R_{\mu \nu }.  \label{L3}
\end{eqnarray}
The action (\ref{action}) does not have a well-defined variational principle, when the
spacetime has a boundary. In order to have a well-defined variational principle, one should add the following
boundary term to the above action
\begin{equation}
I_{\mathrm{b}}=\frac{1}{8\pi }\int_{\partial \mathcal{M}}d^{n}x\sqrt{-h}(K+2\alpha
_{2}J+3\alpha _{3}P),
\end{equation}
where the boundary $\partial \mathcal{M}$ is the hypersurface at some constant $r$, $h_{mn}$ is the induced metric, $K$ is the trace of the extrinsic curvature,
$K_{pq}=\nabla_{(p}n_{q)}$ of the boundary (where the
unit vector $n^{\mu }$ is orthogonal to the boundary and outward-directed)
and $J$ and $P$ are \cite{DM2,Yale}
\begin{eqnarray}
J &=&\frac{1}{3}(3KR-6KR_{mn}R^{mn}+3KK_{mn}K^{mn}-2K_{mn}K^{np}K_{p}^m-K^{3}),  \label{J} \\
P &=& K R^2 - 4K R^{mn}R_{mn} + KR^{mnpq}R_{mnpq} -4K^{mn}R R_{mn} + 12K^{mn}R_m^pR_{np} \nonumber\\
&&- 4K^{mn}R_m^{pqr}R_{npqr} - \frac{2}{3}R K^3 + 4 K^{mn}R_{mn} + 2KK^{mn}K_{mn}R - 12KK^{mn}K_m^pR_{np} \nonumber \\
&&- 4KK^{mn}K^{pq}R_{mpnq} + 12K^{mn}K_m^pK_n^qR_{pq} + 12K^{mn}K_m^pK^{qr}R_{nqpr} - 2K^{mn}K_{mn}K^3 \nonumber \\
&&+ 4K^{mn}K_m^pK_{np}K^2 + 3KK^{mn}K_{mn}K^{pq}K_{pq} - 6KK^{mn}K_m^pK_n^qK_{pq}  \nonumber\\
&&- 4K^{mn}K_{mn}K^{pq}K_p^rK_{qr} + 12K^{mn}R^{pq}R_{mnpq} + \frac{24}{5}K^{mn}K_m^pK_n^qK_p^rK_{qr} + \frac{1}{5}K^5.   \label{P}
\end{eqnarray}

In general, the total action $I_{\mathrm{g}}+I_{\mathrm{b}}$ is not finite when evaluated on the
solution, as is the Hamiltonian and other associated thermodynamic
quantities. In order to have a finite action, we must add some counterterms
to the action (\ref{action}). Here, we restrict ourselves to the case of rotating Lifshitz
black branes with a flat horizon. For this case, as in the case of static solutions of Lovelock
gravity \cite{Deh1}, the following counterterms make the action finite:
\begin{equation}
I_{\mathrm{ct}}=-\frac{1}{8\pi }\int_{\partial \mathcal{M}}d^{n}x\sqrt{-h}\left(%
\frac{(n-1)L^{4}}{l^{5}}+\frac{zq}{2l}\sqrt{-A_{m}A^{m}}\right).
\label{FinAct}
\end{equation}
This is due to the fact that on the boundary $A_{m }A^{m}=-q^{2}$
is constant for rotating Lifshitz solutions. The variation of total action $%
I=I_{\mathrm{g}}+I_{\mathrm{b}}+I_{\mathrm{ct}}$ about a solution of the equation of motion
is
\[
\delta I=\int d^{n}x(S_{mn}\delta h^{mn}+S_{m}\delta A^{m}),
\]
where
\begin{eqnarray*}
S_{mn} &=&\frac{\sqrt{-h}}{16\pi }\left[ \Pi _{mn}+%
\frac{zq}{2l}(-A_{p}A^{p})^{-1/2}(A_{m}A_{n}-A_{p}A^{p}h_{mn})\right] , \\
S_{m} &=&\frac{\sqrt{-h}}{16\pi }[n^{\mu }F_{\mu m}+\frac{zq}{2l}%
(-A_{p}A^{p})^{-1/2}A_{m}],
\end{eqnarray*}
with
\begin{equation}
\Pi _{mn}=K_{mn}-Kh_{mn}+2\alpha_{2}\left( 3J_{mn}-Jh_{mn}\right) +3\alpha _{3}
\left(5P_{mn}-Ph_{mn}\right)
+\frac{(n-1)L^{4}}{l^{5}}h_{mn}. \label{Pi1}
\end{equation}
In Eq. (\ref{Pi1}) $J_{mn}$ and $P_{mn}$ are
\begin{eqnarray}
J_{m n } &=&\frac{1}{3}(2KK_{m p }K_{n }^{p
}+K_{p q }K^{p q }K_{m n }-2K_{m p
}K^{p q }K_{q n }-K^{2}K_{m n }),  \label{Jab} \\
P_{m n } &=&\frac{1}{5}\Big\{[K^{4}-6K^{2}K^{p q
}K_{p q }+8KK_{p q }K_{r }^{q }K^{r
p }-6K_{p q }K^{q r }K_{r s }K^{s
p }+3(K_{p q }K^{p q })^{2}]K_{m n }
\nonumber \\
&&-(4K^{3}-12KK_{r q }K^{r q }+8K_{q r
}K_{s }^{r }K^{s q })K_{m p }K_{n }^{p
}-24KK_{m p }K^{p q }K_{q r }K_{n
}^{r }  \nonumber \\
&&+(12K^{2}-12K_{r s }K^{r s })K_{m p
}K^{p q }K_{q n }+24K_{m p }K^{p q
}K_{q r }K^{r s }K_{n s }\Big\}.  \label{Pab}
\end{eqnarray}

The dual field theory is nonrelativistic for asymptotically Lifshitz
spacetimes, and therefore it will not have a covariant relativistic stress
tensor. However, one can define a stress tensor complex, consisting of the
energy density $\mathcal{E}{}$, energy flux ${}\mathcal{E}_{i}$, momentum
density ${}\mathcal{P}_{i}$, and spatial stress tensor $\mathcal{P}_{ij}$ as
\cite{Ross}
\begin{eqnarray}
\mathcal{E}{}{} &=&2S_{\ t}^{t}-S^{t}A_{t},\quad {}\mathcal{E}{}^{i}=2S_{\
t}^{i}-S^{i}A_{t},  \label{En} \\
{}\mathcal{P}_{i} &=&-2S_{\ i}^{t}+S^{t}A_{i}\quad \mathcal{P}%
_{i}^{j}=-2S_{\ i}^{j}+S^{j}A_{i},  \label{Momen}
\end{eqnarray}
which satisfy the conservation equations
\begin{equation}
\partial _{t}{}\mathcal{E}{}+\partial _{i}{}\mathcal{E}{}^{i}=0,\quad
\partial _{t}{}\mathcal{P}_{j}+\partial _{i}\mathcal{P}_{j}^{i}=0.
\label{Conserv}
\end{equation}

\section{Rotating Lifshitz Black Branes with One Rotation Parameter\label{Rot}}

First, we introduce rotating Lifshitz spacetime in third-order Lovelock
gravity. The Lifshitz metric with one rotation parameter in ($n+1$) dimensions
may be written as \cite{Deh0}
\begin{equation}
ds^{2}=-\frac{r^{2z}}{l^{2z}}(\Xi dt-ad\varphi )^{2}+\frac{l^{2}}{r^{2}}%
dr^{2}+\frac{r^{2}}{l^{4}}(adt-\Xi l^{2}d\varphi
)^{2}+r^{2}\sum_{i=1}^{n-2}dx_{i}{}^{2},  \label{Rot1}
\end{equation}
where $a$ is the rotation parameter and
\[
\Xi ^{2}=1+\frac{a^{2}}{l^{2}}.
\]
The metric (\ref{Rot1}) is a generalization of the asymptotically rotating AdS metric introduced in
\cite{Lem}. Although the metric (\ref{Stat}) may be transformed to the rotating
Lifshitz metric (\ref{Rot1}) by the transformation
\begin{equation}
t\mapsto \Xi t-a\varphi, \ \ \ \ \ \ \ \ \ \ \varphi \mapsto \Xi \varphi -%
\frac{a}{l^{2}}t,  \label{Tr1}
\end{equation}
the metric (\ref{Rot1}) generates a new spacetime if $\varphi $ is
periodically identified. Indeed, the periodic nature of $\varphi $ allows
the metrics (\ref{Stat}) and (\ref{Rot1}) to be locally mapped into each
other but not globally, and so they are distinct \cite{Stach}. We will see
that the rotating solution is not stable for a large rotation parameter, and therefore
the properties of the metrics (\ref{Stat}) and (\ref{Rot1}) are not the same.

Using the ansatz
\begin{equation}
A=q\frac{r^{z}}{l^{z}}\left( \Xi dt-ad\varphi \right) ,
\end{equation}
for the gauge field and defining $\hat{\alpha}_{2}\equiv (n-2)(n-3)\alpha
_{2}$ and $\hat{\alpha}_{3}\equiv (n-2)...(n-5)\alpha _{3}$ for convenience,
it is straightforward to show that the action (\ref{action}) supports the
Lifshitz metric (\ref{Rot1}) provided

\begin{eqnarray}
m^{2} &=&\frac{(n-1)z}{l^{2}},  \nonumber \\
q^{2} &=&\frac{2(z-1)L^{4}}{zl^{4}},  \nonumber \\
\Lambda  &=&-\frac{[(z-1)^{2}+n(z-2)+n^{2}]L^{4}+n(n-1)(\hat{\alpha}%
_{2}l^{2}-2\hat{\alpha}_{3})}{2l^{6}},  \label{Coef5}
\end{eqnarray}
where we define
\begin{equation}
L^{4}\equiv l^{4}-2l^{2}\hat{\alpha}_{2}+3\hat{\alpha}_{3}.  \label{L4}
\end{equation}
for simplicity and we use this definition throughout the paper.

Second, we consider the black brane solution, which asymptotes to the metric (\ref
{Rot1}). The metric of such a solution may be written as
\begin{equation}
ds^{2}=-\frac{r^{2z}}{l^{2z}}f(r)(\Xi dt-ad\varphi )^{2}+\frac{l^{2}}{r^{2}g(r)}%
dr^{2}+\frac{r^{2}}{l^{4}}(adt-\Xi l^{2}d{\varphi })^{2}+r^{2}%
\sum_{i=1}^{n-2}dx_{i}^{2}.  \label{Met2}
\end{equation}
Using the ansatz
\[
A=q\frac{r^{z}}{l^{z}}h(r)\left[ \Xi dt-ad\varphi \right]
\]
for the gauge field, the action (\ref{action}) reduces to
\begin{eqnarray}
I_{\mathrm{g}}&=&\frac{n-1}{16\pi }\int d^{n}xdr\frac{r^{z-1}}{l^{z+1}}%
\sqrt{\frac{f}{g}}\Big\{\left[ r^{n}\left( -\frac{\Lambda }{n(n-1)}l^{2}-g+%
\frac{\hat{\alpha}_{2}}{l^{2}}g^{2}-\frac{\hat{\alpha}_{3}}{l^{4}}%
g^{3}\right)\right]^{\prime } \nonumber \\
&&+\frac{q^{2}r^{n-1}}{2(n-1)f}\left[ (zh+rh^{\prime })^{2}g+m^{2}l^{2}h^{2}%
\right] \Big\}.  \label{Act2}
\end{eqnarray}

Functionally varying (\ref{Act2}) with respect to $f(r)$, $g(r)$, and $h(r)$
yields upon simplification
\begin{eqnarray}
&&\hat{\alpha}_{3}\left[ (n+6z-6)f+3rf^{\prime {}}\right] g^{3}+\hat{\alpha}%
_{2}l^{2}\left[ (n+4z-4)f+2rf^{\prime {}}\right] g^{2}  \nonumber \\
&&-l^{4}\left[ (n+2z-2)f+rf^{\prime {}}\right] g-\frac{\Lambda l^{6}}{n-1}f=%
\frac{q^{2}l^{4}}{2(n-1)}\left[ (zh+rh^{\prime })^{2}g -m^{2}l^{2}h^{2}\right],  \nonumber\\
&&\left\{ r^{n}\left( -\frac{\Lambda l^{2}}{n(n-1)}-g+\frac{\hat{\alpha}_{2}}{%
l^{2}}g^{2}-\frac{\hat{\alpha}_{3}}{l^{4}}g^{3}\right) \right\} ^{\prime }
=\frac{q^{2}r^{n-1}}{2(n-1)f}\left[ (zh+rh^{\prime })^{2}g+m^{2}l^{2}h^{2}%
\right],  \nonumber \\
&&2r^{2}h^{\prime {\prime {}}}-r\left[ (lnf)^{\prime {}}-(lng)^{\prime {}})%
\right] (rh^{\prime {}}+zh)+2(z+n)rh^{\prime {}}+2(n-1)zh=2m^{2}l^{2}\frac{h%
}{g}.  \label{Eqs}
\end{eqnarray}
The numerical solutions of the above field equations (\ref{Eqs}) are the
same as those given in \cite{Deh0}. This is due to the fact that the
rotating metric (\ref{Met2}) and static metric ($\Xi =1$) are locally the
same and therefore the metric functions are the same in both cases.

\section{Thermodynamics of Rotating Lifshitz Black Branes\label{Therm}}

Now, we investigate the thermodynamics of rotating black branes. The
temperature of the event horizon is given by
\begin{equation}
T=\frac{1}{2\pi }\left( -\frac{1}{2}\nabla _{b}\xi
_{a}\nabla ^{b}\xi ^{a}\right) _{r=r_{0}}^{1/2},  \label{Tem1}
\end{equation}
where $\xi $ is the Killing vector
\begin{equation}
\xi =\partial _{t}+\Omega \partial _{\varphi }
\end{equation}
and $\Omega $ is the angular velocity of the Killing horizon given as
\begin{equation}
\Omega =-\left[ \frac{g_{t\varphi }}{g_{\varphi \varphi }}\right] _{r=r_{0}}=\frac{%
\sqrt{\Xi ^{2}-1}}{\Xi l}.  \label{Om}
\end{equation}
Using Eq. (\ref{Tem1}) and the expansion of the metric functions given in
the appendix (see Eq. (\ref{ExpH})), the temperature can be obtained as
\begin{equation}
T=\frac{r_{0}^{z+1}}{4\pi l^{z+1}\Xi }\left( f^{\prime }g^{\prime }\right)
_{r=r_{0}}=\frac{r_{0}^{z+1}\sqrt{f_{1}g_{1}}}{4\pi l^{z+1}\Xi }.
\label{Temp}
\end{equation}
The form of the near-horizon expansion given in the appendix suggests that $%
f_{1}$ and $g_{1}$ are proportional to $r_{0}^{-1}$, and therefore
\begin{equation}
T=\frac{\eta }{4\pi \Xi }r_{0}^{z},  \label{T3}
\end{equation}
where $\eta $ is a proportionality constant.

The entropy can be calculated through the use of \cite{Myers2}
\begin{equation}
S=\frac{1}{4}\sum_{i=1}^{p}k\alpha _{i}\int d^{n-1}x\sqrt{\tilde{g}}\tilde{%
\mathcal{L}}_{i-1},  \nonumber
\end{equation}
where the integration is done on the ($n-1$)-dimensional spacelike
hypersurface of the Killing horizon with induced metric $\tilde{g}_{ab}$
(whose determinant is $\tilde{g}$), and $\tilde{\mathcal{L}}_{i}$ is the $i$%
th order Lovelock Lagrangian of $\tilde{g}_{ab}$. Since we are dealing with
flat horizon, $\tilde{\mathcal{L}}_{1}=\tilde{\mathcal{L}}_{2}=0$ and
therefore the entropy density is
\begin{equation}
S=\frac{1}{4}\Xi r_{0}^{n-1}.  \label{Ent}
\end{equation}

The conserved quantity along the radial coordinate $r$, which relates to the
coefficients of the expansion of the metric function at the horizon ($r=r_{0}
$) and at infinity is \cite{Deh1}
\begin{equation}
\mathcal{C}_{0}=\{(1-2\frac{\hat{\alpha}_{2}}{l^{2}}g+3\frac{\hat{\alpha}_{3}%
}{l^{4}}g^{2})[rf^{\prime }+2(z-1)f]-q^{2}(zh+rh^{\prime })h\}\frac{r^{n+z-1}%
}{l^{z+1}}(\frac{f}{g})^{1/2}.  \label{C0}
\end{equation}
Using the expansion of metric functions at the horizon given in the appendix
and Eqs. (\ref{Temp}-\ref{Ent}), the constant $\mathcal{C}_{0}$ (\ref
{C0}) at $r=r_{0}$ can be written as
\begin{equation}
\mathcal{C}_{0}=\frac{r_{0}^{n+z}\sqrt{f_{1}g_{1}}}{l^{z+1}}=16\pi TS=\eta
r_{0}^{n+z-1}.  \label{C0hor}
\end{equation}
\ On the other side, by use of the large $r$ expansions given in the
appendix for the metric functions, we obtain
\begin{equation}
\mathcal{C}_{0}=\digamma C_{1},  \label{C0inf}
\end{equation}
where
\begin{equation}
\digamma =\frac{2(z-1)(z+n-1)^{2}\{(z-n+1)l^{4}+2(n-2)\hat{\alpha}%
_{2}l^{2}-3(n+z-3)\hat{\alpha}_{3}\}}{zl^{z+5}\mathcal{K}}  \label{Flov}
\end{equation}
for Lovelock gravity and
\begin{equation}
\digamma =\frac{4(n-1)}{(2n-3)l^{n}}  \label{Fein}
\end{equation}
in the case of Einstein gravity with $z=n-1$.

Using Eqs. (\ref{En}) and (\ref{Momen}), one can compute the energy and
angular momentum densities of the black brane for $z\neq n-1$ as
\begin{eqnarray}
\mathcal{E} &=&\frac{(n+z-1)\Xi ^{2}-z}{16\pi (z+n-1)}\digamma C_{1}=\frac{%
(n+z-1)\Xi ^{2}-z}{16\pi (z+n-1)}\eta r_{0}^{n+z-1},  \label{Et} \\
J &=&\frac{l}{16\pi }\Xi \sqrt{\Xi ^{2}-1}\digamma C_{1}=\frac{l}{%
16\pi }\Xi \sqrt{\Xi ^{2}-1}\eta r_{0}^{n+z-1},  \label{Jt}
\end{eqnarray}
where $\digamma $ is given in Eq. (\ref{Flov}). The above expressions are
valid for the case of $z=n-1$ provided $\digamma $ reads from Eq. (\ref{Fein}%
). One may note that Eq. (\ref{Et}) reduces to that given in \cite{Deh1} for
the case of the static Lifshitz solution ($\Xi =1$), and the angular momentum
density vanishes as expected.

Now, using Eqs. (\ref{C0hor}), (\ref{C0inf}), (\ref{Et}) and (\ref{Jt}), we
find
\begin{equation}
\mathcal{E}=\frac{n-1}{n-1+z}TS+\Omega J.  \label{flot}
\end{equation}
Again for $\Xi =1$, Eq. (\ref{flot}) reduces to that of static case since
both $\Omega $ and $J$ vanish. Also, it is worth noting that
Eq. (\ref{flot}) reduces to
\begin{equation}
\mathcal{E}=\frac{n-1}{n}TS+\Omega J.  \label{AdS}
\end{equation}
for asymptotic AdS black branes \cite{DKh}. Recently, it has been shown that the
Komar-conserved quantity corresponding to the null Killing vector is also proportional to the Hawking temperature \cite{Ban}.

Dividing Eq. (\ref{Et}) by Eq. (\ref{Jt}) gives us a Smarr-type formula as
\[
\mathcal{E}(S,J)=\frac{(n+z-1)\Xi ^{2}-z}{(z+n-1)l\Xi \sqrt{\Xi
^{2}-1}}J,
\]
where $\Xi $ depends on $S$ and $J$ through the following
equation
\[
16\pi J-\eta l\left( \frac{4S}{\Xi }\right) ^{(n+z-1)/(n-1)}\Xi
\sqrt{\Xi ^{2}-1}=0.
\]
One may then regard the parameters $S$ and $J$ as a complete set
of extensive parameters for the energy density $\mathcal{E}(S,J)$
and define the intensive parameters conjugate to $S$ and $J$.
These quantities are the temperature and the angular velocity
\begin{equation}
T=\left( \frac{\partial \mathcal{E}}{\partial S}\right) _{J},\quad \Omega
=\left( \frac{\partial \mathcal{E}}{\partial J}\right) _{S}.
\label{Inten}
\end{equation}
It is a matter of straightforward calculation to show that the intensive
quantities calculated by Eq. (\ref{Inten}) coincide with Eqs. (\ref{Om}) and
(\ref{T3}) found in Sec. \ref{Therm}. Thus, the thermodynamic quantities
calculated in this paper satisfy the first law of thermodynamics
\[
d\mathcal{E}=TdS+\Omega dJ.
\]

\section{Stability of Rotating Lifshitz black branes}

Now, we investigate the stability of rotating Lifshitz black branes. The
stability of a thermodynamic system with respect to the small variations of
the thermodynamic coordinates can be carried out by finding the determinant
of the Hessian matrix of the energy with respect to its extensive parameters
$X_{i}$, $\mathbf{H}_{X_{i}X_{j}}^{\mathcal{E}}=[\partial ^{2}\mathcal{E}%
/\partial X_{i}\partial X_{j}]$. In our case the energy density is a
function of the entropy and angular momenta densities. The number of
thermodynamic variables depends on the ensemble which is used. In the
canonical ensemble, the angular momenta is a fixed parameter, and therefore
the positivity of heat capacity, $C_{J}=T(\partial T/\partial S)_{J}$, is
sufficient to assure the local stability. This quantity can be calculated as
\[
C_{J}=\frac{S\left[ (n-z-1)\Xi ^{2}+z\right] }{(n+2z-1)(\Xi ^{2}-1)+z\Xi ^{2}%
}.
\]
Since the denominator is positive, the heat capacity is positive provided \
\begin{equation}
(n-z-1)\Xi ^{2}+z>0.  \label{Cond}
\end{equation}
In the grand-canonical ensemble, the determinant of the Hessian matrix of
the energy with respect to $S$ and $J$ can be obtained as
\[
\left| \mathbf{H}_{S,J}^{\mathcal{E}}\right| =\frac{z}{\Xi ^{2}l^{2}S^{2}%
\left[ (n-z-1)\Xi ^{2}+z\right] },
\]
which shows that the Lifshitz black branes are stable provided the condition
(\ref{Cond}) holds.

Thus, the condition of stability of Lifshitz black branes is the same in
both canonical and grand-canonical ensembles. One may note that the
stability analysis is also the same in Einstein and Lovelock gravity for
Lifshitz black branes with a flat boundary. For $z\leq n-1$, the condition (%
\ref{Cond}) is satisfied and therefore Lifshitz black branes are stable.
But, they can be unstable for $z>n-1$ provided the rotation parameter $a$
satisfies
\begin{equation}
a>\left( \frac{n-1}{z-(n-1)}\right) ^{1/2}l.  \label{Con-a}
\end{equation}

\section{CONCLUSIONS}

In this paper, we investigated the thermodynamics of rotating Lifshitz black
branes in Lovelock gravity. We obtained the conserved quantities of the
solutions through the use of counterterm method introduced in \cite{Deh1}.
We showed that the energy density $\mathcal{E}$, entropy density $S$,
temperature $T$, angular velocity $\Omega $ and angular momentum density $J$
are related by Eq. (\ref{flot}). This relation is the generalization of
Eq. (\ref{AdS}) for asymptotic AdS black branes. Also, one may note that
the relation (\ref{flot}) reduces to the relation between the energy density
$\mathcal{E}$, entropy density $S$, and temperature $T$ for static Lifshitz
solutions. We also obtain a Smarr-type formula for the energy density $%
\mathcal{E}(S,J)$ as a function of extensive quantities $S$ and $J$. We
showed that the thermodynamic quantities calculated in this paper satisfy
the first law of thermodynamics. Finally, we investigated the stability of
the black brane solutions in both canonical and grand-canonical ensemble. We
found that the solutions are thermally stable for the solutions with $z\leq
n-1$, while they can be unstable for $z>n-1$ provided the rotation parameter
satisfies Eq. (\ref{Con-a}). It is worth noting that, for $z>2(n-1)$, the
rotation parameter can be less than $l$.

One may generalize rotating Lifshitz black brane solutions with one rotating
parameter to the case of rotating solutions with more rotation parameters.
The rotation group in $n+1$ dimensions is $SO(n)$ and therefore the number
of independent rotation parameters is [$n/2$], where [$x$] is the integer
part of $x$. Equation (\ref{flot}) for Lifshitz black branes with $k\leq
\lbrack n/2]$ rotation parameters generalizes to
\[
\mathcal{E}=\frac{n-1}{n-1+z}TS+\sum\limits_{i=1}^{k}\Omega _{i}J_{i},
\]
where $\Omega _{i}$ and $J_{i}$'s are the $i$th component of angular
velocity and angular momentum, respectively. Also, one may generalize the
above equation to the case of charged rotating Lifshitz black branes in the presence of Maxwell field.

\section{Appendix}

In this appendix, we consider the near-horizon and large-$r$ behavior of
rotating Lifshitz solutions. Since we consider nonextreme black brane
solutions, the functions $f(r)$ and $g(r)$ go to zero linearly, that is
\begin{eqnarray}
&&f(r)=f_{1}(r-r_{0})+f_{2}(r-r_{0})^{2}+f_{3}(r-r_{0})^{3}+f_{4}(r-r_{0})^{4}+...
\nonumber \\
&&g(r)=g_{1}(r-r_{0})+g_{2}(r-r_{0})^{2}+g_{3}(r-r_{0})^{3}+g_{4}(r-r_{0})^{4}+...
\nonumber \\
&&h(r)=f_{1}^{1/2}%
\{h_{1}(r-r_{0})+h_{2}(r-r_{0})^{2}+h_{3}(r-r_{0})^{3}+h_{4}(r-r_{0})^{4}+..%
\},  \label{ExpH}
\end{eqnarray}
where $f_i$'s, $g_i$'s and $h_i$'s are constants which can be
obtained through the use of field equations \cite{Deh1}.

In order to have the appropriate asymptotic behavior for the metric
functions, one may use the straightforward perturbation theory:
\begin{eqnarray*}
&&f(r)=1+\epsilon w_{f}(r), \\
&&g(r)=1+\epsilon w_{g}(r), \\
&&h(r)=1+\epsilon w_{h}(r),
\end{eqnarray*}
where $\epsilon$ is an infinitesimal parameter. It is a matter of calculations to show that the metric functions at large $r$
are \cite{Deh1}
\begin{eqnarray}
h(r) &=&1-\epsilon \left\{ \frac{C_{1}}{r^{n+z-1}}+\frac{C_{2}}{%
r^{(n+z-1+\gamma )/2}}+\frac{C_{3}\Theta (n+z-1-\gamma )}{r^{(n+z-1-\gamma
)/2}}\right\} ,  \nonumber \\
f(r) &=&1-\epsilon \left\{\frac{C_{1}F_{1}}{r^{n+z-1}}+\frac{C_{2}F_{2}}{r^{(n+z-1+\gamma
)/2}}+\frac{C_{3}F_{3}\Theta (n+z-1-\gamma )}{r^{(n+z-1-\gamma )/2}}\right\},
\nonumber \\
g(r) &=&1-\epsilon \left\{\frac{C_{1}G_{1}}{r^{n+z-1}}+\frac{C_{2}G_{2}}{r^{(n+z-1+\gamma
)/2}}+\frac{C_{3}G_{3}\Theta (n+z-1-\gamma )}{r^{(n+z-1-\gamma )/2}}\right\},
\label{Finf}
\end{eqnarray}
where $C_i$'s are arbitrary constants,
\begin{eqnarray*}
\gamma &=&\left\{ (17-8\mathcal{B})z^{2}-2(3n+9-8\mathcal{B})z+n^{2}+6n+1-8%
\mathcal{B}\right\} ^{1/2}, \\
F_{1} &=&2\left( z-1\right) \left( z-n+1\right) {\mathcal{K}}^{-1}, \\
F_{2} &=&\left( \mathcal{F}_{1}-\mathcal{F}_{2}\right) \left\{ 8z\mathcal{K}%
\left[ (z-1)\mathcal{B}+2n+z-3\right] \right\} ^{-1}, \\
F_{3} &=&\left( \mathcal{F}_{1}+\mathcal{F}_{2}\right) \left\{ 8z\mathcal{K}%
\left[ (z-1)\mathcal{B}+2n+z-3\right] \right\} ^{-1}, \\
G_{1} &=&2\left( z-1\right) \left( n+z-1\right) {\mathcal{K}}^{-1}, \\
G_{2} &=&\left( \mathcal{G}_{1}+\mathcal{G}_{2}\right) \left\{ 8z\mathcal{K}%
\left[ (z-1)\mathcal{B}+2n+z-3\right] \right\} ^{-1}, \\
G_{3} &=&\left( \mathcal{G}_{1}-\mathcal{G}_{2}\right) \left\{ 8z\mathcal{K}%
\left[ (z-1)\mathcal{B}+2n+z-3\right] \right\} ^{-1}, \\
\mathcal{K} &=&(z-1)(n+z-1)\mathcal{B}+z(z-1)+n(n-1), \\
\mathcal{F}_{1} &=&8(z-1)[(z-1)(n+z-1)\mathcal{B}+z(z-1)+n(n-1)] \\
&&\times \lbrack (z-1)(n+3z-3)\mathcal{B}-2z^{2}+(n+3)z+n(n-2)-1], \\
\mathcal{F}_{2} &=&\gamma \lbrack n-1+(z-1)\mathcal{B}]\Big\{8(1+\mathcal{B}%
)(z-1)^{3} \\
&&+(17n-9+8\mathcal{B})(z-1)^{2}+2(n+8)(n-1)(z-1)+n^{2}(n-1)-(n-1)\gamma ^{2}%
\Big\}, \\
\mathcal{G}_{1} &=&8(z-1)[2(z-1)\mathcal{B}-3z+3n-1][(z-1)(n+z-1)\mathcal{B}%
+z(z-1)+n(n-1)], \\
\mathcal{G}_{2} &=&\Big\{8(1+\mathcal{B})(z-1)^{3}+(17n-9+8\mathcal{B}%
)(z-1)^{2} \\
&&+2(n+8)(n-1)(z-1)+n^{2}(n-1)\Big\}\gamma -(n-1)\gamma ^{3},
\end{eqnarray*}
$\mathcal{B}=(l^{4}-4\hat{\alpha}_{2}l^{2}+9\hat{\alpha}_{3})/L^{4}$, and
\[
\Theta (n+z-1-\gamma )=\left\{
\begin{array}{cc}
1 & n+\,\,z-1>\gamma \hspace{.5cm}(z<\frac{n-\mathcal{B}}{2-\mathcal{B}}) \\
&  \\
0 & n+\,\,z-1\leq \gamma \hspace{.5cm}(z\geq \frac{n-\mathcal{B}}{2-\mathcal{%
B}})
\end{array}
\right.
\]

In the case of $z=n-1$ in Einstein gravity, the metric functions at large $r$
are \cite{Deh1}
\begin{eqnarray}
h(r) &=&1-\epsilon \frac{C_{1}\ln r+C_{2}}{r^{2(n-1)}},  \nonumber \\
f(r) &=&1-\epsilon \frac{3n-4}{(n-1)(2n-3)}\frac{C_{1}}{r^{2(n-1)}},
\nonumber \\
g(r) &=&1-\epsilon \left\{ \frac{2(n-2)(C_{1}\ln r+C_{2})}{(2n-3)r^{2(n-1)}}+%
\frac{(n^{2}-2)C_{1}}{(n-1)(2n-3)^{2}r^{2(n-1)}}\right\} .  \label{FinfSpz}
\end{eqnarray}

\acknowledgements
This work has been supported by the Research Institute for Astrophysics and
Astronomy of Maragha.

\end{document}